\UseRawInputEncoding
\documentclass[preprint, aps, prd, superscriptaddress, nofootinbib, floatfix]{revtex4-2}

\usepackage[T1]{fontenc}
\usepackage[utf8]{inputenc}
\usepackage{amsmath, amssymb, physics, bm}
\usepackage{graphicx, xcolor}
\usepackage{siunitx}
\usepackage{ulem}
\usepackage{booktabs}
\usepackage{comment}
\usepackage{stackengine}
\usepackage{adjustbox}
\usepackage{multirow}
\usepackage{caption}
\usepackage{geometry}
\usepackage{accents}
\usepackage{trimclip}

\newcommand\ringring[1]{%
  {\mathop{\kern0pt #1}\limits^{
    \vbox to-1.85ex{
      \kern-2ex
      \hbox to 0pt{\hss\normalfont\kern.1em \r{}\kern-.45em \r{}\hss}
      \vss
    }
  }}
}
\usepackage[colorlinks=true, linkcolor=blue, citecolor=blue, urlcolor=blue]{hyperref}
\usepackage{cleveref}

\begin{document}

\flushbottom % important for better vertical spacing control

\title{\Large The Noisy Universe}

\author{Gabriela Barenboim}
\affiliation{Instituto de F\'{i}sica Corpuscular, CSIC-Universitat de Val\`{e}ncia, Paterna 46980, Spain\\[-0.1ex]}
\affiliation{Departament de F\'{i}sica Te\`{o}rica, Universitat de Val\`{e}ncia, Burjassot 46100, Spain\\[-0.1ex]}

\author{Aurora Ireland}
\affiliation{Department of Physics, University of Chicago, Chicago, IL 60637, USA\\[-0.1ex]}
\affiliation{Leinweber Institute for Theoretical Physics, Stanford University, Stanford, CA 94305, USA\\[-0.1ex]}

\author{Albert Stebbins}
\affiliation{Fermi National Accelerator Laboratory, Theoretical Astrophysics Group, Batavia, IL 60510, USA\\[-0.1ex]}

\email{gabriela.barenboim@uv.es}
\email{anireland@stanford.edu}
\email{stebbins@fnal.gov}

\begin{abstract}
We present observational constraints on large-scale white noise (LSWN) in the cosmic density field, a phenomenon predicted to arise from non-linear mode coupling during cosmological evolution. Building on the theoretical framework of Paper I, where we demonstrated that non-linearities inevitably redistribute power from small to large scales through mode mixing, we confront these predictions with current cosmological data. We modify the CLASS Boltzmann code to incorporate a white noise component $k_\mathrm{BH}/k$ in the primordial power spectrum and perform parameter estimation using current cosmological data. 
The non-detection of excess power on the largest observable scales places stringent upper bounds: $k_\mathrm{BH} \leq 1.80 \times 10^{-13}~\mathrm{Mpc}^{-1}$ at 99\% confidence. These constraints imply the primordial power spectrum must deviate from 
 a power law on small scales, perhaps with a sharp cutoff at 
$k_{\mathrm{cut}} \lesssim 0.03~\mathrm{pc}^{-1}$
or through running of the spectral index with $\alpha_s \lesssim -0.019$.
Our results demonstrate that LSWN provides a powerful probe of the primordial spectrum at scales orders of magnitude smaller than those directly observable, offering unique constraints on early-universe physics.
\end{abstract}

\maketitle

\pagebreak

%%%%%%%%%%%%%%%%%%%%%%%%%%%%%%%%%%%%%%%%%%%%%%%%%%%%%%%%%%%%
%%%%%%%%%%%%%%%%%%%%%%%%%%%%%%%%%%%%%%%%%%%%%%%%%%%%%%%%%%%%
\section{Introduction}\label{sec:introduction}

The equations of motion (EoMs) which describe the evolution of cosmic inhomogeneities are spatially homogeneous partial differential equations (PDEs).  It is well known that non-linearities in such PDEs couple different spatial scales (wavenumbers), whereas for linear PDEs different spatial scales evolve independently.  In early universe cosmology, the inhomogeneities are extremely small, $\sim10^{-5}$, and non-linear terms in the EoMs are even smaller, $\sim10^{-10}$.  However, this does not mean that non-linearities are unimportant or that they cannot dominate on some spatial scales (wavenumbers).  
With the appropriate parameterization of cosmic inhomogeneities --- parameterizations which obey PDEs --- the inhomogeneities on very large super-horizon scales are very small. Consequently, non-linear mode coupling between very small and very large scales can rapidly ``spill power'' from small to large scales and generate the dominant contribution to the power spectrum on the largest scales.\footnote{In the language of scattering, such large scale spillage can be described by $\vec{q}_1+\vec{q}_2\rightarrow\vec{q}$ when  $|\vec{q}|\ll|\vec{q}_1|,\,|\vec{q}_2|$.}  This power generically has a white noise spectrum
 and contributes no super-horizon correlations. We refer to this phenomenon as large-scale white noise (LSWN), and it is described in both general and detailed terms in Paper I~\cite{PaperI}.

While mode coupling of cosmic inhomogeneities has been considered previously, this generic LSWN phenomenon has not been previously recognized, perhaps for a variety of reasons: 1) the scale of the relevant non-linearities and the scale where LSWN manifests are typically separated by 9 orders of magnitude, which makes it hard to discern;
2) lore about conserved quantities --- while mode coupling to large scales is suppressed for conserved quantities, the quantities which describe curvature inhomogeneities have no corresponding conservation laws to prevent the generation of LSWN;
3) lore about causality --- while causality forbids the generation of spatial correlations on super-horizon scales, such correlations are not required for LSWN. Indeed, by definition, pure white noise has zero spatial correlations.

In this paper we search for evidence of LSWN in cosmological measurements and discuss implications of a negative result.

\subsection{Relic LSWN and Cosmic Confusion}

The case most relevant for cosmology is one in which the non-linearities responsible for producing most of the LSWN occur on very small scales in the distant past.  This LSWN we call ``relic'' LSWN since it is generated early on and is insensitive to non-linearities at more recent times.  Relic LSWN should already be present late in the radiation era and early in the matter era, while the inhomogeneities are still small.  The late-time large-scale relic power spectrum for locally measurable perturbation variables is
\begin{equation}\label{eq:RelicPowerSpectrum}
    P[t,k]\approx(P_\mathrm{prim}[k]+P_\mathrm{LSWN})\,|T[t,k]|^2
\end{equation}
where $P_\mathrm{prim}[k]$ is the primordial super-horizon power spectrum, $P_\mathrm{LSWN}$ is the LSWN amplitude, and $T[t,k]$ is the linear theory transfer function at late times.
Since both contributions share the same transfer function
$|T(t,k)|^2$, their ratio is time-independent; the dominance of LSWN at
small $k$ is a consequence of spectral shape alone --- $P_\mathrm{prim}
\propto k^{n_s}$ (sub-Poissonian, for $n_s>0$) versus $P_\mathrm{LSWN}
\propto k^0$ --- and holds regardless of the overall amplitude.

The perturbation variable must be locally measurable so that its EoM is a PDE\footnote{The comoving curvature perturbation $\cal R$ is not locally measurable nor is its non-linear EoM a PDE.}.  An example of such a variable is the kurvature density $\Delta\rho=\rho-\theta^2/(24\pi G)
=\nabla^2\mathcal{R}/(4\pi G)$ defined in Paper I~\cite{PaperI} and Paper {\large $\alpha$}~\cite{Stebbins2025}.

An unfortunate characteristic of the relic LSWN power spectrum,
Eq.~(\ref{eq:RelicPowerSpectrum}), is that if one only views the power spectrum on the large scales where it is valid, one could absorb $P_\mathrm{LSWN}$ into $P_\mathrm{prim}[k]$, \textit{i.e.}~a white noise component of the primordial power spectrum is indistinguishable from white noise generated by non-linearities.  This we call ``cosmic confusion'', which limits what one can infer about small-scale non-linearities from the large-scale power spectrum.

\subsection{LOBA and the No-No-Scale Theorem}

There are a variety of phenomena which could occur on small scales and produce non-linear couplings to curvature inhomogeneities and hence LSWN: gravitational waves, phase transitions, etc. The minimal contribution comes from the non-linear self-coupling of the inhomogeneities with themselves.  In Paper I, it was shown how to model this with the Leading Order\footnote{The leading-order non-linearity for density inhomogeneities is quadratic.} Born Approximation (LOBA), from which one obtains a LSWN amplitude of the form
\begin{equation}
    P_\mathrm{LSWN}\approx \int d^3\vec{q}\,(W[|\vec{q}|]\,P_\mathrm{prim}[|\vec{q}|])^2 \,,
\end{equation} 
where $W[q]$ is a weight function which depends on the non-linearities in the hydrodynamics of the cosmological fluid at wavenumber $q$. 
This integral is the $k\to0$ limit of a more general expression for arbitrary wavenumber.  The "whiteness" in LSWN is merely the statement that this $k\to0$ limit is finite, neither zero nor infinite.  The full shape of the power spectrum generated by non-linearities is governed by the wavenumbers which contribute the dominant non-linearities. For conventional cosmological models the dominant non-linearities occur at very early times and at very small scales and are not accessible observationally. The power spectrum of non-linearity-generated inhomogeneities is effectively flat (white) over the entire observable range.
Clearly $W[k]$ and $P_\mathrm{prim}[k]$ cannot both be a power law for $P_\mathrm{LSWN}<\infty$, \textit{i.e.}~either the matter dynamics or the primordial power spectrum (or both) must define a characteristic scale.  This we call the No-No-Scale Theorem. For relic LSWN, convergence at the high-$k$ end requires a high-$q$ cutoff in either $W$ or $P_\mathrm{prim}$ (or both).  
 $W[q]$ is determined by hydrodynamical/gravitational physics and cannot be changed without changing the matter properties. Varying the model dependent $P_\mathrm{prim}[q]$ seems a more likely proposition.

It is easy to understand how non-linearities can generate LSWN which comes to dominate the power spectrum during radiation domination.  In this era, inhomogeneities grow while outside the sound horizon and then oscillate inside the sound horizon. The Fourier transform of any non-linear function of inhomogeneities which has power concentrated at large $k$ will go to a constant as $k \rightarrow 0$, \textit{i.e}~non-linearities have LSWN. It is the smallest $k$'s, which spend the most time growing outside the horizon, which come to dominate the large $k$ inhomogeneities that generate the white noise. Another way of stating this is to note that the sound horizon is the Jeans length and the cosmological fluid is gravitationally unstable for wavelengths longer than the Jeans length.  Any process that generates inhomogeneities with these long wavelengths will trigger the inherent instability of the cosmological fluid.  The acoustic waves themselves do just that.

During the radiation era the linear evolution of every sinusoidal wave is just a scaled version of every other wave. Because of this scale-free\footnote{Photon/neutrino viscosity does break scale invariance but only mildly.} radiation era dynamics $W[q]$ is accurately a power law in $q$ for wavenumbers which cross the horizon during this era. For a pure power law $P_\mathrm{prim}[q]$ the No-No-Scale theorem then limits the range of wavenumbers which enter the horizon during the radiation era and for $n_\mathrm{s}\approx1$ this cutoff must come at large $q$, \textit{i.e.}~a late beginning of radiation domination. Another possibility is that the power law is cutoff at large $q$.  In either case, measuring or putting limits on $P_\mathrm{LSWN}$ tells us something about small scales and early times, \textit{e.g.}~the comoving horizon scale during the beginning of radiation domination or the cutoff scale for the primordial power spectrum (or both).

\section{Parameter Constraints}
\label{sec:Constraints}

The main quantitative result of this paper is to constrain the amplitude of LSWN in our universe.  Current cosmic microwave background (CMB) and large-scale structure measurements show a nearly scale-invariant spectrum with amplitude $A_s \approx 2.1 \times 10^{-9}$ and spectral index $n_s \approx 0.965$ extending to scales $k \sim 0.1~\text{Gpc}^{-1}$, with no evidence for deviation from a power law and thus no evidence for LSWN~\cite{Planck:2019nip}. Thus we cannot expect to measure the LSWN and our conclusions will concern implications of \textit{not} detecting LSWN on the largest observable scales. 

A preliminary constraint on the amplitude of LSWN  was given in Paper I.  This was an order-of-magnitude estimate based on the idea the LSWN cannot dominate on observable scales, which extend up to $\sim10\,\mathrm{Gpc}$.  The new constraints presented here are based on modern parameter estimation techniques. We base this analysis on a fiducial flat $\Lambda$CDM cosmological model characterized by six primary parameters: the physical baryon density $\Omega_b h^2$, the physical cold dark matter density $\Omega_c h^2$, the angular scale of the sound horizon at recombination $\theta_\star$, the optical depth to reionization $\tau$, the amplitude of primordial scalar perturbations $A_s$, and their spectral index $n_s$. We assume a spatially flat universe with $\Omega_\mathrm{tot} = 1$, neglect primordial tensor perturbations, and fix the sum of neutrino masses to $\sum m_\nu = 0.059\ \mathrm{eV}$, consistent with a normal mass ordering.

We compute numerical simulations of the evolution of the CMB and matter power spectrum, evolving curvature perturbations through radiation and matter eras using a modified CLASS Boltzmann code~\cite{Lesgourgues:2011he}. Parameter estimation is carried out using the COBAYA Monte Carlo Markov Chain (MCMC) framework~\cite{Torrado:2020dgo}, applying a convergence threshold of $R - 1 < 0.01$ based on the Gelman-Rubin statistic. The fit utilizes the Planck 2018 TT, TE, and EE angular power spectra~\cite{Planck:2019nip}, Planck lensing data~\cite{Planck:2018lbu}, and the Atacama Cosmology Telescope (ACT) Data Release 6 lensing likelihood~\cite{ACT:2023kun}. Additional constraints on large-scale structure are provided by Baryon Acoustic Oscillation (BAO) measurements from the DESI DR2 dataset~\cite{DESI:2025dr2II}.

As demonstrated in Paper I and stated in Eq.~(\ref{eq:RelicPowerSpectrum}), the super-horizon power spectrum of curvature perturbations can be written as the primordial power spectrum with an additional relic white noise component. At the linear level, in the literature, the curvature power spectrum is expressed as $\Delta^2_\mathcal{R}[k]=(4\pi)^3 G^2 P[k]/k$, where $P[k]$ is the power spectrum of $\Delta\rho$.  Thus Eq.~(\ref{eq:RelicPowerSpectrum}) is
\begin{equation}
    \Delta^2_\mathcal{R}[k] \approx {}_{(1)}\Delta^2_\mathcal{R}[k] +\frac{k_\mathrm{BH}}{k} \,,
\end{equation}
where ${}_{(1)}\Delta^2_\mathcal{R}[k]=(4\pi)^3 G^2 P_\mathrm{prim}[k]/k$ and
$k_\mathrm{BH}=(4\pi)^3 G^2 P_\mathrm{LSWN}$, the wavenumber defined by $\Delta_\mathcal{R}^2[k_{\rm BH}] \approx 1$, parametrizes the amplitude of LSWN. Our parametric cosmological model includes all the standard parameters of $\Lambda$CDM plus $k_\mathrm{BH}$.  Specifically, we model the primordial power spectrum as

\begin{equation}
\Delta^2_\mathcal{R}[k] = A_\mathrm{s}\,
\left( \frac{k}{k_0} \right)^{n_\mathrm{s} -1} +  \frac{k_\mathrm{BH}}{k}
=
A_\mathrm{s}\,
\left(\frac{k}{k_0}\right)^{n_\mathrm{s}-1}\left(
1+\left(\frac{k_\mathrm{LSWN}}{k}\right)^{n_\mathrm{s}}
\right)\,,   
\end{equation}
where $k_0=0.05\,\text{Mpc}^{-1}$ is the CMB pivot scale. The wavenumber where the primordial power equals the LSWN is $k_\mathrm{LSWN}=(
k_\mathrm{BH}\,k_0^{1-n_\mathrm{s}}/A_\mathrm{s}
)^{1/n_\mathrm{s}}$.
Since $n_\mathrm{s}>0$, any LSWN enhances power on large spatial scales (small $k$).  This should be most apparent in the low multipoles of the CMB.

Since the leading order non-linearities are quadratic, in the scenario where there is a sharp cutoff in the primordial spectrum at $k=k_\mathrm{cut}$, the correction due to non-linearities is roughly
${}_{(2)}\Delta_\mathcal{R}^2[k_\mathrm{cut}]\sim
({}_{(1)}\Delta_\mathcal{R}^2[k_\mathrm{cut}])^2$.  Since the relic kurvature spectrum is white on super-horizon scales significantly larger than $1/k_\mathrm{cut}$\footnote{The whiteness of the spectrum is not a long timescale dynamical effect nor an extrapolation but simply the 
asymptotic form in $k$ space of any non-linearity.  This spectrum is not modified on super-horizon scales where the growth 
factor is $k$-independent.}, 
i.e.  $P_K\propto k^0$, ${}_{(2)}\Delta_\mathcal{R}^2[k]\propto1/k$, we may very roughly approximate 
$(4\pi)^3\,G^2\,P_\mathrm{LSWN}\sim k_\mathrm{cut}\,({}_{(1)}\Delta_\mathcal{R}^2[k_\mathrm{cut}])^2$ so
\begin{equation}
    k_\mathrm{BH} \sim A_\mathrm{s}^2\, \left(\frac{k_\mathrm{cut}}{k_0}
          \right)^{2\,(n_\mathrm{s}-1)} k_\mathrm{cut}
\qquad\qquad
k_\mathrm{LSWN}\sim A_\mathrm{s}^{\frac{1}{n_\mathrm{s}}}
\left(\frac{k_\mathrm{cut}}{k_0}
      \right)^{\frac{n_\mathrm{s}-1}{n_\mathrm{s}}}
k_\mathrm{cut}
    \,.
\end{equation}
In this truncated power law model, since $n_\mathrm{s}\approx1$, LSWN is comparable to the primordial power only at wavenumbers $\lesssim A_\mathrm{s}\sim10^{-10}$ times smaller than the scales which generate the LSWN! In this sense, when the primordial inhomogeneities are small, $A_\mathrm{s}\ll1$, LSWN is a very subtle effect.

Alternatively, if the primordial spectrum exhibits running of the spectral index, $n_\mathrm{s}[k] = n_\mathrm{s}[k_0] + \alpha_\mathrm{s}\,\ln[k/k_0]$, a sufficiently negative running $\alpha_\mathrm{s}$ will suppress LSWN by gradually reducing power toward small scales. Both scenarios --- cutoff and running --- provide mechanisms to limit LSWN generation, and observational constraints on $k_\mathrm{BH}$ thus constrain the small-scale behavior of the primordial spectrum.

\begin{figure}
    \centering
    \includegraphics[width=0.9\linewidth]{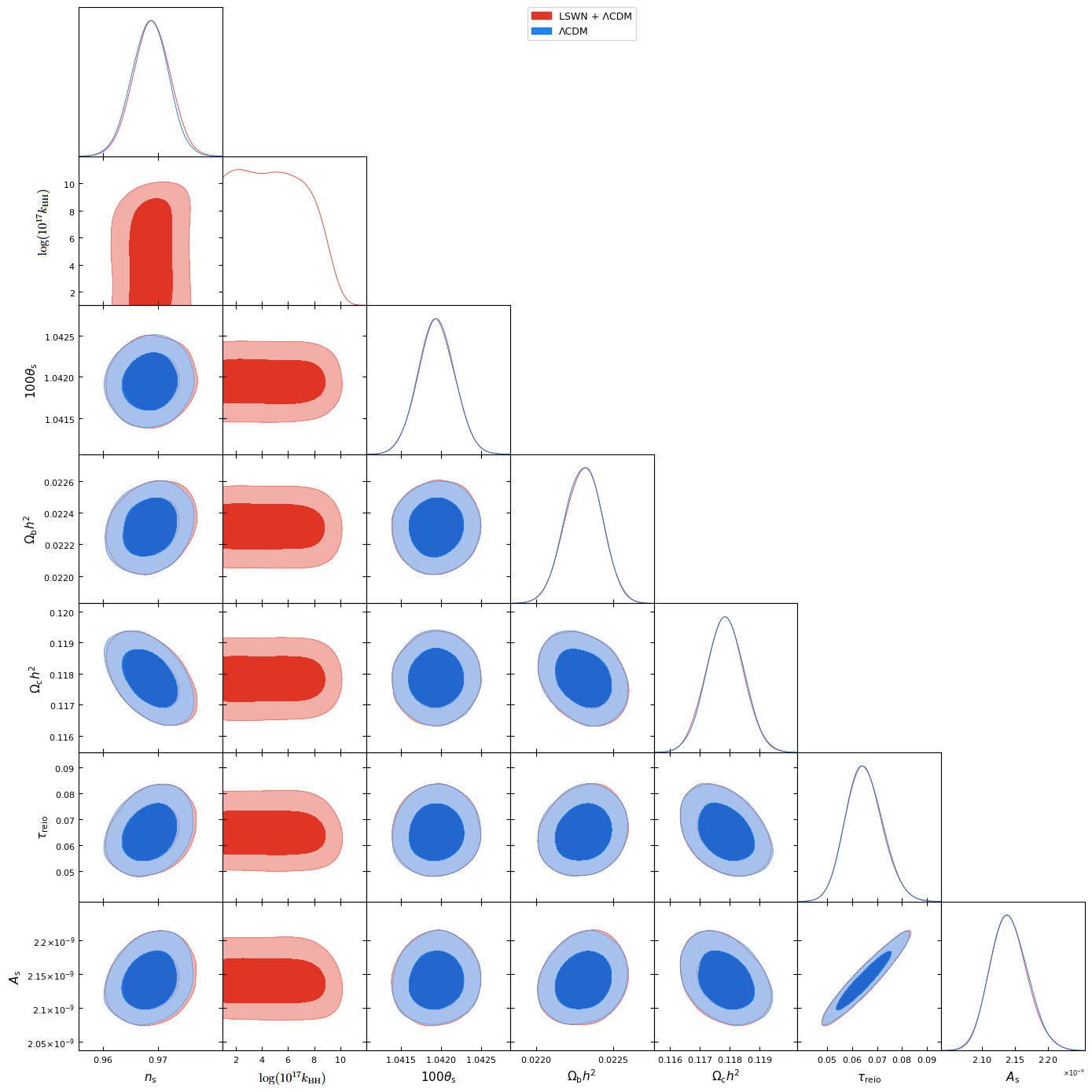}
   \caption{Constraints on the 6-parameter base $\Lambda$CDM model (blue), and the 7-parameter model including the LSWN coefficient $k_\mathrm{BH}$ (red). The regions lie almost exactly on top of each other, confirming that $k_\mathrm{BH}$ is uncorrelated with standard cosmological parameters.}
    \label{fig:triangleplot}
\end{figure}

In Fig.~\ref{fig:triangleplot} we constrain our model with and without the addition of the $k_\mathrm{BH}$ parameter.  The difference between these two cases on the standard parameters is negligible and barely discernible in this figure. We find that there is no evident parameter degeneracy of $k_\mathrm{LSWN}$ with other parameters and that the constraints on the other parameters are not significantly loosened by the introduction of the additional parameter.  Allowing the universe to also have spatial curvature does not significantly change the constraint.

Our main result is an upper bound on the amplitude of LSWN:
\begin{eqnarray}
k_\mathrm{BH} \leq 9.94\times 10^{-14} \;\mbox{Mpc}^{-1}\;\;\mbox{at $95 \%$ C.L.}
\nonumber \\
k_\mathrm{BH} \leq 1.80 \times 10^{-13}  \;\mbox{Mpc}^{-1} \;\; \mbox{at $99\%$ C.L.}
\end{eqnarray}
These observational limits have profound implications for the small-scale primordial power spectrum. Using the scaling relation above with fiducial values $A_s \approx 2.1 \times 10^{-9}$ and $n_s \approx 0.965$, the constraint $k_\mathrm{BH} \leq 10^{-13}$ Mpc$^{-1}$ implies:

\begin{itemize}
\item \textbf{Cutoff model}: The primordial spectrum must be cut off at comoving scales  $k_\mathrm{cut} \lesssim 0.03~\mathrm{pc}^{-1}$, corresponding to physical wavelengths greater than about 30 parsec so LSWN is predominantly generated when the cosmic temperature is below $\sim3\,\mathrm{MeV}$.  This cutoff scale is  five orders of magnitude smaller than directly observable scales ($\sim 10$ Mpc), yet remains far above microphysical scales.

\item \textbf{Running model}: Alternatively, running of the spectral index with 
 $\alpha_\mathrm{s} \lesssim -0.019$ suffices to suppress LSWN below observable levels. Unlike constraints on $A_\mathrm{s}$ and $n_\mathrm{s}$, the LSWN constraint \textit{requires} non-zero running or a cutoff—the primordial spectrum cannot be a pure power law extending to arbitrarily small scales.

\end{itemize}

\noindent The constraint is driven primarily by CMB measurements at low multipoles ($\ell \lesssim 30$), where white noise contributes as $C_\ell \propto \ell(\ell+1)\,k_\mathrm{BH}/k$ and is least diluted by primordial power. The non-detection of excess power on these largest observable scales thus places stringent limits on non-linear processes operating at scales far below direct observational reach.

\section{Implications}\label{sec:implications}

The question we address is not whether LSWN exists --- Paper I established its inevitability --- but rather how large it is in our universe and what this tells us about the small-scale primordial spectrum, particularly whether future observations can detect or constrain the predicted running or cutoff. The answer will either confirm that inflation generated nearly scale-invariant fluctuations down    to scales of tens of parsec, or reveal new physics operating at these intermediate scales. Either outcome would represent a significant advance in our understanding of the early universe.

Defining $k_\mathrm{LSWN}$ as the wavenumber where $P_\mathrm{LSWN}=P_\mathrm{prim}[k_\mathrm{LSWN}]$, \textit{i.e.}~where LSWN starts to dominate, in the cutoff model we find that if $n_\mathrm{s}\approx1$, then $k_\mathrm{LSWN}\approx\sqrt{k_\mathrm{cut} k_\mathrm{BH}}$.  The three scales, $k_\mathrm{cut},\,k_\mathrm{LSWN},\,k_\mathrm{BH}$, each differ by a factor of $\sim A_\mathrm{s}$, or 9 orders of magnitude.  Our upper limit $k_\mathrm{BH}\lesssim10^{-13}\,\mathrm{Mpc}^{-1}$ thus corresponds to $k_\mathrm{LSWN}\lesssim0.1\,\mathrm{Gpc}^{-1}$, which was the value estimated in Paper I.  Thus, our preliminary constraints  $k_{\rm cut} \lesssim 0.03~\text{pc}^{-1}$ and $\alpha_s \lesssim -0.019$  are confirmed by a more detailed comparison with data.  
Direct observational constraints on the power spectrum start to become uncertain below $\sim10\,\mathrm{Mpc}$. The requirement of a cutoff at a scale 
greater than $\sim30\mathrm{pc}$ is telling us about inhomogeneities on scales  $\sim5$ orders of magnitude smaller than what can be probed directly.  This constraint exists because if the o below $\sim30\,\mathrm{pc}$, non-linearities from these sound waves crossing the horizon would have generated larger LSWN than is allowed.

A less dramatic modification of $\Lambda$CDM that would satisfy observational constraints is to introduce a small negative running of the spectral index: $n_\mathrm{s}[k] = n_\mathrm{s}[k_0] + \alpha_\mathrm{s}\,\ln[k/k_0]$ with $\alpha_\mathrm{s} \lesssim -0.015$. While this is consistent with current Planck uncertainties ($\alpha_s = -0.006 \pm 0.013$)~\cite{Planck:2019nip}, it also represents a falsifiable prediction. Future measurements could detect the required $\alpha_\mathrm{s}$.

Whether through a cutoff or running, production of LSWN from non-linearities in the cosmological fluid \textit{requires} a modification of the primordial power spectrum from that used in the $\Lambda$CDM model.  Furthermore, self coupling of non-linearities from passively evolving acoustic waves is only the minimal constraint. As pointed out in Paper {\large $\alpha$}, {\it any} process generating shear in the matter flow will couple non-linearly to the curvature and contribute to the generation of LSWN.  Acoustic oscillations are only one such process. First-order phase transitions, gravitational waves, topological defects, or other exotic matter all contribute to the matter shear and will produce additional LSWN.  Constraints from LSWN on these other processes will be given in future work.

\section{Summary}\label{sec:summary}

In this Letter, we have argued that non-linearities in the evolution of cosmological inhomogeneities will inevitably generate large scale white noise (LSWN) which will dominate the inhomogeneity power spectrum at the longest wavelengths.  Properly accounting for this requires the introduction of a new parameter $k_\mathrm{BH}$, which parametrizes the amplitude of LSWN. Comparison with current cosmological data constrains this parameter to be smaller than $k_\mathrm{BH}\leq 1.80 \times 10^{-13}~\mathrm{Mpc}^{-1}$ (99\% CL).  The implication of this constraint is that, at minimum, the observed power law power spectrum must be cut off at small scales. If the cutoff is sharp, then this scale is at or larger than 
 about 30 comoving parsec.  A smoother cutoff is also a possibility.  For example, a negative running of the spectral index  $\alpha_\mathrm{s}\lesssim-0.019$ would satisfy the constraint from LSWN.  In either case, a primordial power law power spectrum is no longer viable over a dynamic range greater than $\sim 9$ orders of magnitude in wavelength,  from $30\,\mathrm{pc}$ to $10\,\mathrm{Gpc}$ comoving. This limitation on the power spectrum will have profound implications for models for the generation of cosmic inhomogeneities. This will be explored in future work.

Furthermore, constraints on the power spectrum given here are the \textit{minimal} constraints.  Other processes can contribute to LSWN which would tighten constraints on the power spectrum and also put limits on phenomena on small scales in the early universe, such as phase transitions, gravitational waves, etc.  These constraints will also be explored in future work.

Perhaps most fundamentally, our work demonstrates that in spite of the fact that cosmological inhomogeneities are very very small, non-linearities on the smallest scales can and do inevitably lead to large modifications of the inhomogeneities on large scales at late times. This finding challenges the conventional wisdom that linear theory suffices for analyzing large-scale structure. Our results establish LSWN as both a theoretical necessity and an observational target, opening a new frontier in precision cosmology and providing an accessible empirical probe of phenomena on the smallest scales.

\section*{Acknowledgements}

GB is supported by the Spanish grants   CIPROM/2021/054 (Generalitat Valenciana), PID2020-113775GB-I00 
(AEI/10.13039/501100011033), and by the European ITN project HIDDeN (H2020-MSCA-ITN-2019/860881-HIDDeN). AI is supported by NSF Grant PHY-2310429, Simons Investigator Award No.~824870, DOE HEP QuantISED award \#100495, the Gordon and Betty Moore Foundation Grant GBMF7946, and the U.S.~Department of Energy (DOE), Office of Science, National Quantum Information Science Research Centers, Superconducting Quantum Materials and Systems Center (SQMS) under contract No.~DEAC02-07CH11359. AS is supported by FermiForward Discovery Group, LLC. This manuscript has been authored by FermiForward Discovery Group, LLC under Contract No. 89243024CSC000002 with the U.S. Department of Energy, Office of Science, Office of High Energy Physics.

\bibliographystyle{apsrev4-2} 
\bibliography{Noisy}

\end{document}